# Has been Neutrinoless Double Beta Decay of $^{130}$Te observed by the CUORE experiment ?
Comments to the paper arXiv:1504.02454v1 [nucl-ex] 9 Apr 2015

I.V. Kirpichnikov
SSC RF "Institute for Theoretical and Experimental Physics", Moscow
NRC "Kurchatov Institute"

A recent reconsideration of a theoretical process of a neutrinoless double beta decay [1,2] points out that a signal of the decay should be shifted by a few keV from the Q-value for the 2ß0ν-decay. The conclusion [1,2] was based on an analysis of the experimental data of the HEIDELBERG-MOSCOW experiment [3] and a reconsideration of the Cuoricino data [4,5].

Still the recent CUORE results [6] contradicted the conclusions of [1,2] completely. No trace of a 2β0ν-decay was found and only a limit $T_{1/2} > 2.8 \cdot 10^{24}$ y was postulated. The problem was very serious and the CUORE results have deserved a more careful consideration.

**The CUORE results.**

The CUORE-0 detector has parameters very close to those of the first presentation by E.Fiorini (Neutrino-2004 [4]) and Cuoricino [ Physical Review C 78, 035502 (2008) [5] ]. The only "innovation" was introducing of the data "salting" within the energies near the expected 2ß0ν-decay peak (a kind of a blinding). *And it seems that this procedure killed the effect.*

The detector contained 52x750g bolometers, the active mass of Te-130 was about 11 kg ($TeO_2$ : 39 kg). The recent CUORE paper (arXiv:1504.02454v1, nucl-ex) demonstrated a progress of the project. The background was improved successfully. A resolution of the device was 5.2 keV compare to 7.2 keV in the previous Cuoricino measurements.

The most serious improvement concerned a background level : it was three times lower near the $Q_{\beta\beta}$-value of $^{130}$Te then at Cuoricino-2008 data , and close to the E.Fiorini 2004 result. The collected statistics was Mt=9.8 kg y   ( 2004 y – 5.8 kg y,  2008 y – 11.83 kg y ).

No peaks were found in the energy interval E = (2470-2568) keV except for the sum Co-60. A flat <bcgnd> was accepted as <N>=4.0 events/2 keV.  The $\chi2$ – value for the above energy interval was $\chi2$ =43.9/46=0.954  (Fig1).

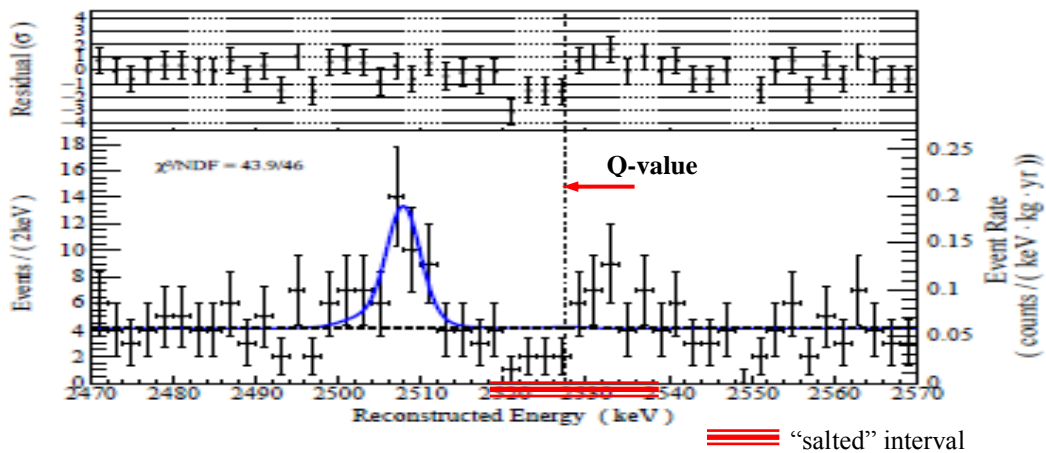

Fig.1.

The treatment of the data was perfect. But a visible structure of the spectrum within the "salted" energy interval (2518-2544) keV was ignored.  If one compared three separated energy intervals , results would be quite different :

| ΔE = | (2470-2996)keV | (2542-2568)keV | (2518-2544)keV |
|---|---|---|---|
| $\chi2$ = | 26/14/4= =0.464 | 45/14/4= =0.804 | 73/14/4= = 1.304 |

The $\chi2$ – value for the "salted" interval (2518-2544) keV was two times higher then the mean value for the two others $\chi2$ =71/27/4=0.656 .

A comparison with the Cuoricino spectrum of 2004 y (Mt=5.29 kg y [4b]) confirmed a distortion of the CUORE spectrum (fig.2 and Table #1).

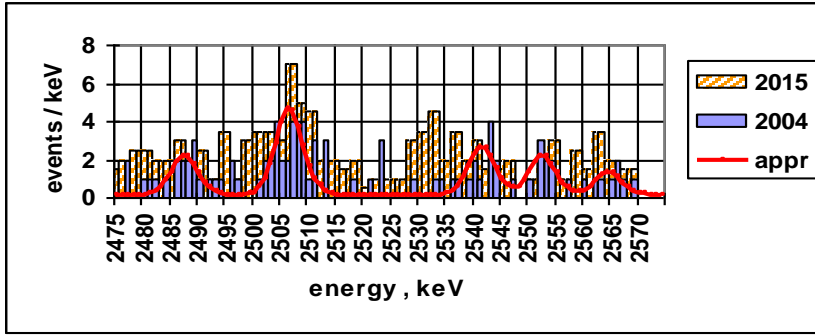

Peaks in the 2004 y spectrum were positioned rather arbitrary (all half-widths were 7.2 keV ). A flat component was attributed to 2615 keV peak.

fig.2

Table #1. Distributions of events over different energy intervals.

| Energy → | 2470-2484 | 2518-2527 | 2528-2537 | 2538-2554 | 2555-2564 | Except 2528-37 |
|---|---|---|---|---|---|---|
| ΔE , keV | 15 | 10 | 10 | 17 | 10 | 52 |
| 2004(Mt=5.29 kg y) | 10 | 5 | 5 | 19 | 9 | 43 |
| 2015(Mt=9.8 kg y) | 30 | 11 | 35 | 29 | 31 | 101 |
| 2015/04 | 3.0 | 2.2 | 7.0 | 1.5 | 3.4 | <2.34> |

**The 2528-2537 keV interval had much higher 2015/04 ratio !**

The same conclusion could be made from a comparison of the CUORE results with the Cuoricino data of 2008 y ( Cuoricino , Mt=11.83 kg y ; CUORE, Mt=9.8 kg y).

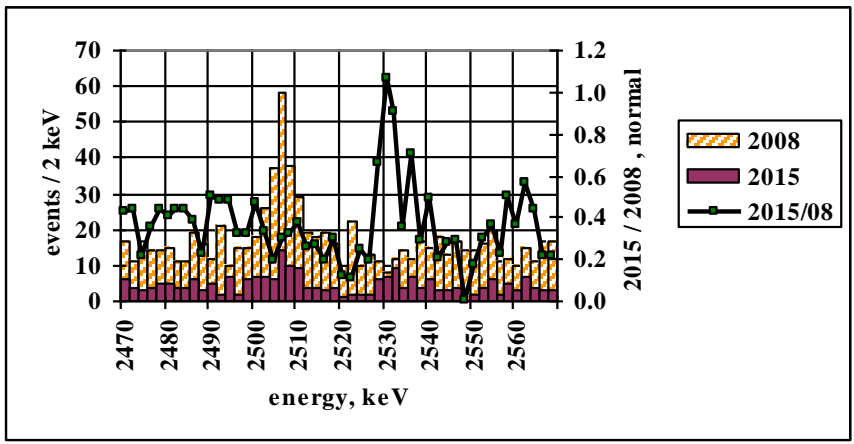

Green points showed ratios of numbers of events in the spectra (the right scale, normalized).
A sharp maximum between 2528 and 2542 kev could be explained, of course, by an existence of an unknown background peak in the 2015 y spectrum.

Fig.3

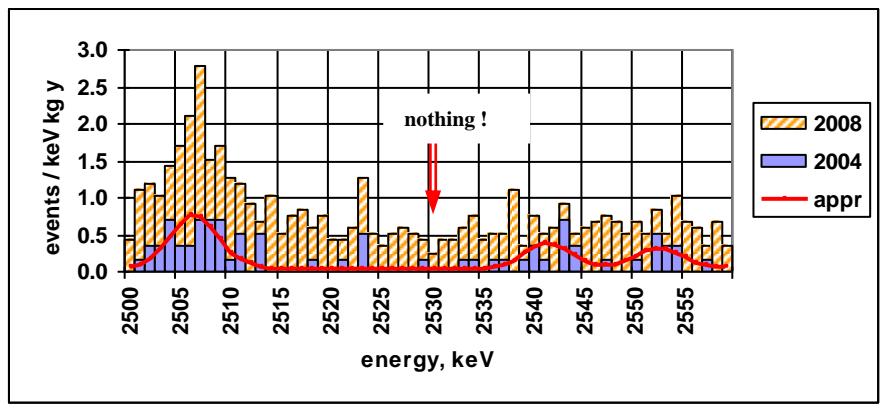

But no trace of such a peak was observed in both the 2004 y or 2008 y spectra (fig.4). Therefore it was quite reasonable to connect a visible distortion of the spectrum just with a procedure of the "salting".

Fig.4

**It was supposed that all events due to the 2β0ν-effect were shifted by several keV to higher energies, and the total number of events in the spectrum was conserved. It could be used for an estimation of the possible effect of 2β0ν-decay.**

**A choice of a background I.**

The estimation demanded a proper choice of the CUORE background. First, the background was compared with Compton gammas from a 2615 keV peak [6a]. It was found that a very simple approximation (Fig.5) provided a possibility to reproduce the CUORE spectrum just near Q-value (fig.6). The same approximation was able to reproduce the spectrum within the total energy interval, which has been presented in the original paper (fig.7,8). So it was supposed that the experimental spectrum could be and should be compared just with the Compton gammas from a 2615 keV peak

**2015 data and a gamma-spectrum due to the 2615 keV line**
**( arXiv:1504.02454v1 [nucl-ex] 9.04.15 – fig.1 )**

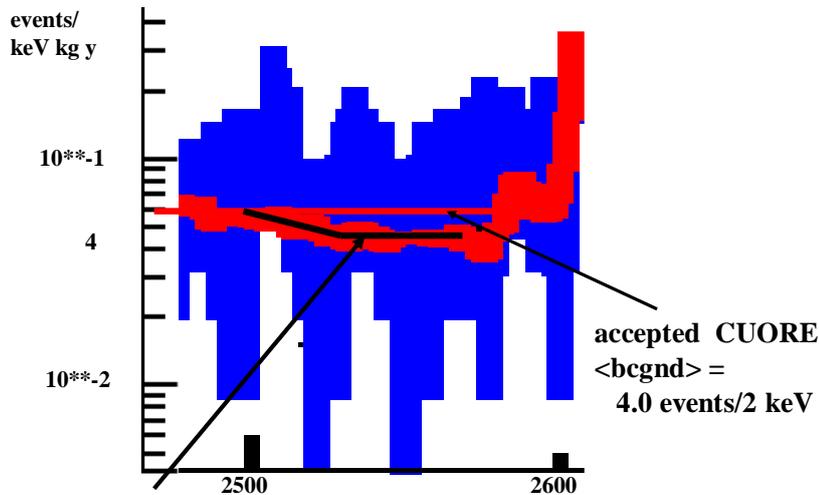

Fig.5

**The estimated gamma <bcgnd> for the ROI due to 2614 MeV peak = 2.3 events/2 keV**

A background value <n>=2.3 for the ROI was found by a postulation (Fig.6):

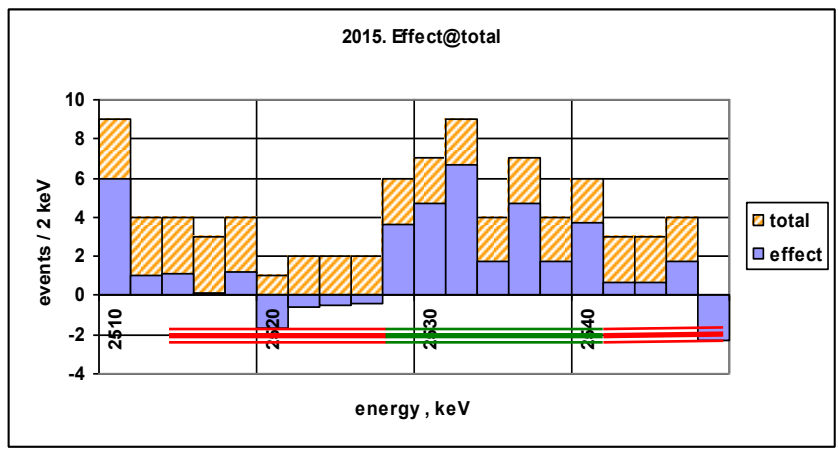

(∑n/11-2.3)=0 for side energies, where ∑n=∑events within energy intervals [(2514-2528)+(2542-2548)] keV.
  Extra events were calculated within an energy region (2528-2440) keV.
   An additional number of events was ΔN=26.9,
or ΔN = 2.7 events / kg y.

Fig.6.

**A choice of the CUORE background II.**

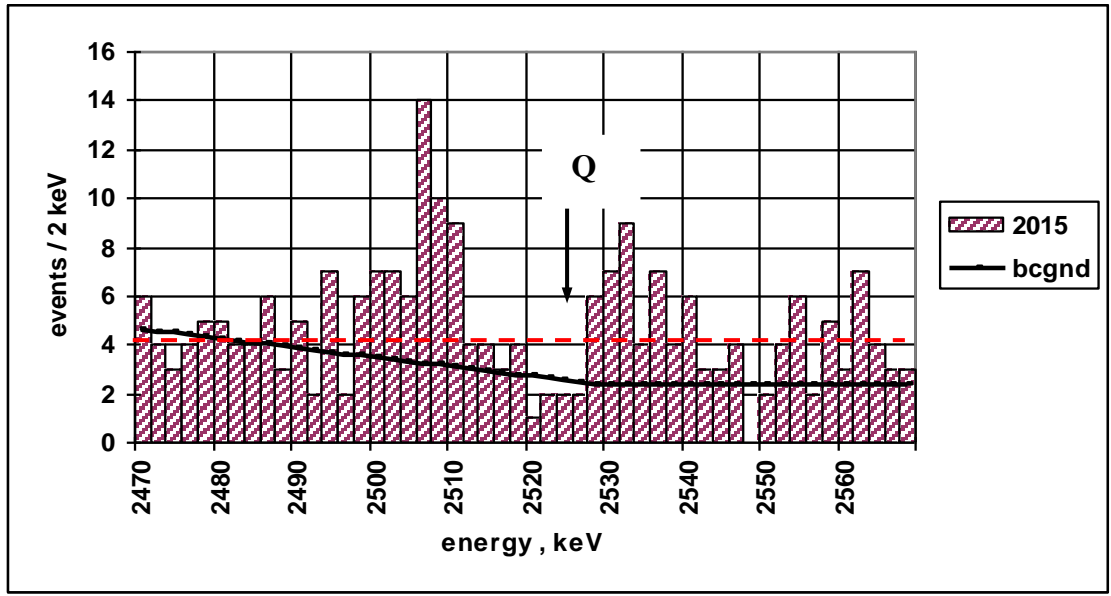

fig.7

The estimated gamma-background due to the 2615 line was chosen according to data of the paper arXiv:1504.02454v1[nucl-ex]9Apr2015 (the black solid line; see also fig.5). The dashed line showed the background accepted in the CUORE paper. A structure at E=(2550-2568) keV was indicated also in the 2004 y data (fig.2).

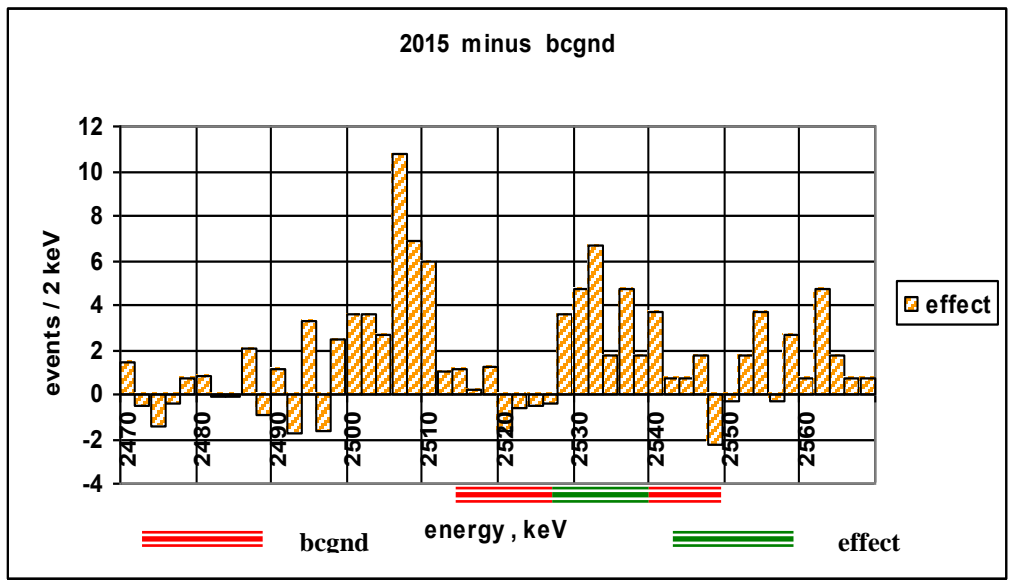

fig.8.

The gamma-background (N=2.3 events / 2 keV within the ROI) was subtracted.
Extra events were calculated in the energy interval ΔE=(2528-2540) keV.
An additional number of events was ΔN=25.0, or ΔN = 2.7 events / kg y.

The number of extra events was exactly the same as it was given in the [2]. So one could accept these events as a shifted signal of 2β0ν-decay of $^{130}$Te. It meant that the CUORE data didn't contradict the conclusions of [1,2], but, on the contrary, supported them.

It was also supported by a comparison of Cuoricino 2008 y and CUORE data .

## A comparison of Cuoricino 2008 and CUORE 2015

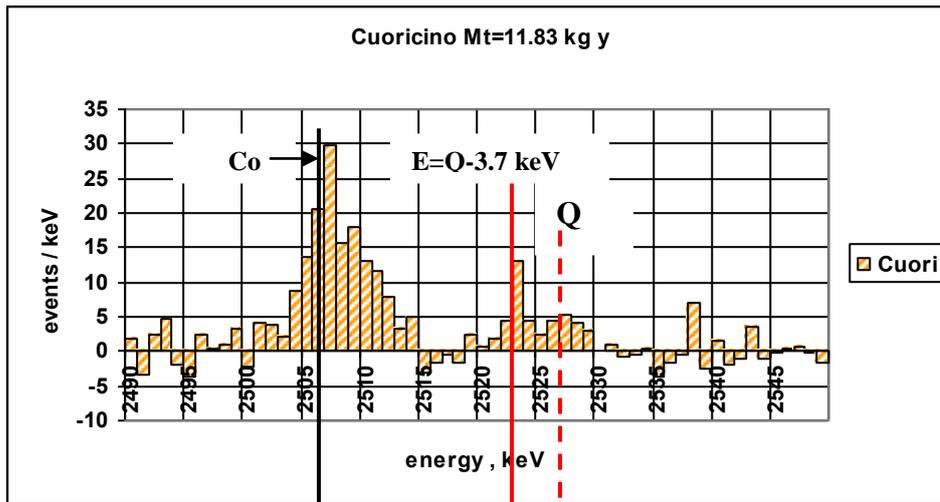

fig.9

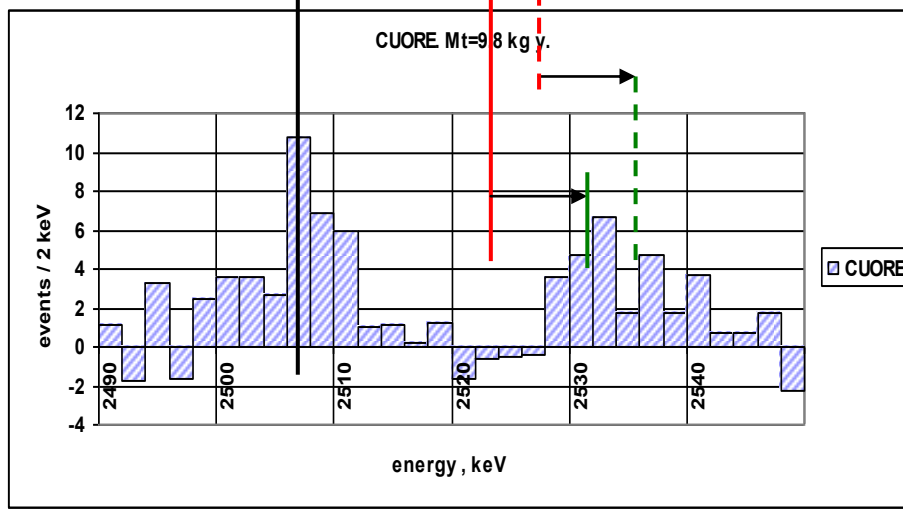

fig.10.

Positions of the sum Co peaks were the same. The unknown line in the CUORE spectrum indicated presence of the two components, both of which were shifted by ~8 keV relative to the Cuoricino data.

### Why one could see the shifts.

The product nucleus of the 2β0ν-decay process was born in the ground state. The proper product atom was in an excited state. The excitation could be estimated as a difference ΔE between sum energies of electrons in the proper shells of a parent and a product atom, hence ΔE being about several keV. This excitation could be removed only by emission of X-rays, as an emission of electrons was energy disfavored. The responses of Ge and bolometric Xe detectors for X-rays are different, so signals of the 2β0ν-decay would be different also.

### Ge-detector.

A bunch of X-rays was immediately captured in Ge–detector, producing a bunch of low-energy electrons. This "suite" always accompanied new-born electrons of 2β0ν-decay. It provided a distinction between shapes of the signals, which were produced by a capture of electrons of 2β0ν-decay, and electrons with the same energies from Compton scattering of gammas. The signal of 2β0ν-decay contained a more intensive long component.

Elements of a spectrometric tract produced a kind of a pulse shape discrimination, which resulted in a very weak shift of the signal, some about $10^{-3}$ of the electron energies. The shift depended on parameters of a spectrometric tract, and it could explain an observed slightly different shifts of the peak in the data of H-M Collaboration and GERDA.

Te-bolometer.

The quite different process was responsible for a visible shift of the 2β0ν-decay in the Te-bolometric detector. The shifted signal could be seen only if a part of the X-rays bunch left the detector. And it was possible due to a $TeO_2$ crystal structure, which provided a possibility for a canalization of X-rays [2].

The total signal of the detector should include both the components, the shifted and non-shifted ones. A presence of the two peaks were clearly indicated in the Cuoricino data (fig.9 ). The result of CUORE supported the claim (fig.10). It confirmed a particular nature of the line and could be accepted as a proof for the existence of the 2β0ν-decay.

**A conclusion**

The new model of a neutrinoless 2β-decay [1,2] predicted shifts of 2β0ν-signals from the $Q_{\beta\beta}$ values . It changed the situation with a search for 2β0ν-decay completely : according to the author of [1,2] the process was observed ten years ago in the two experiments.

The model was based on an analysis of the experimental data of the HEIDELBERG-MOSCOW experiment [3] and a reconsideration of the Cuoricino results [4] . It has been shown that the recent CUORE data have not contradicted the conclusions of [1,2] and , on the contrary , have supported them. A life-time of 2β0ν-decay of $^{130}$Te could be estimated from the CUORE data as $T_{1/2} \approx 1 \cdot 10^{-24}$ years in the full agreement with the author of [2]

**Finally, 2β0ν-decay of $^{76}$Ge and $^{130}$Te was observed ten years ago in two experiments.**